TRE    (05.05.2014.)

# Time and the river of existence

Mario Radovan
University of Rijeka,
Department of informatics,
51000 Rijeka,
Croatia

mradovan@inf.uniri.hr
http://www.inf.uniri.hr/~mradovan/

**Abstract**

Time is one of those issues about which many thinkers and scientists have tried to pronounce their finest thoughts, but the discourse about time has remained vague and often inconsistent. In this paper we put forward a conceptual framework inside which the issue of time should be addressed and solved. We argue that time is an abstract entity created by the human mind, not an ingredient of physical reality as physics normally assumes. Physical reality is a process of ceaseless becoming and vanishing; time is not a part of that process. Time does not flow: time is the artificial bank in relation to which we measure the intensity and amount of the flow (change) of physical reality. It is necessary to differentiate physical reality and abstract entities by means of which we describe this reality. It is necessary to differentiate formulas from their interpretations: a correct formula can be interpreted in a logically inconsistent and factually wrong way.

## 1. The river and its bank

"Time is one of the last great mysteries", runs the first sentence of a large anthology of texts about time (Callender, 1). Since the ancient times, many have tried to explain what time is, but it seems that nobody has done this in a quite successful way. Physics produced various images of time, which are not mutually compatible and some of them seem inconsistent. Time has been called a river that carries forward every thing until it sinks in its waves, but time is not a river and it does not flow. Physical reality is a process and it can be called the *river of existence*. This river does not need time or any other force beyond itself to "carry" it and to be what it is: a *process*. Time is not part of that process: it is not an ingredient of physical reality and it does not flow. Time is the *artificial bank* in relation to which the river of physical reality flows. Time is an element of our language by means of which we *speak* about physical reality and its change: it is one of the basic elements of the conceptual system by means of which we express our perception and understanding of the physical reality in which we dwell and of which we are part (Radovan 2011).

Time is an abstract dimension on which the human mind projects its experience of the changing reality. Aging and death are not brought about by time, but by the change which is immanent to physical world. The passage of time and our passage in time are metaphorical expressions of the fact that everything changes. This common metaphor has its poetical beauty and charm, but taken literally, it is wrong.

We define and measure time by means of various cyclic processes, but a cyclic process is not time: it is only a process. Time is an abstract measure of the amount and intensity of change, expressed in terms of some chosen cyclic process, such as the rotation of the earth around the sun and around itself, the oscillation of certain atoms or other particles. Physical reality is a process of becoming and vanishing; the human mind projects its perception and experience of this process to an abstract dimension called time. Points on this dimension are determined by a cyclic process which is the most stable we currently know, or sufficiently stable for certain purposes. In sum, time is a creation of the human mind: it is a metaphysical (linguistic, cognitive) category, not a part of the physical world. The same holds for other basic elements of the conceptual system by means of which we express our experience and understanding of existence. Other core elements of that system are space and causality; numbers must be added to the list of core metaphysical categories in terms of which we think and speak.

In the next section we put forwards a matrix of basic concepts which must be understood correctly for a clear discourse about time; this matrix consists of three columns, each of which has three elements. The first column speaks about three kinds of reality: (1) physical, (2) mental, and (3) abstract. The second column regards knowledge; its elements are: (1) reality in itself, (2) formal descriptions, and (3) the interpretations of formal descriptions. The third column describes three levels of discourse: (1) ontology, (2) epistemology, and (3) logic. For a precise and coherent discourse about anything, it is necessary to understand these nine concepts (Radovan 2013).

## 2. The matrix of discourse

We often encounter questions such as "Does time exist?" and "Is time real or unreal?". Those who ask such questions do not tell us what it means to exist, to be real or unreal. Let us introduce an ontological framework in which every entity belongs into one of the following three classes: (1) *physical entities*, (2) *mental entities*, and (3) *abstract entities*. Let us label the class (space or world) of physical entities with W1, the class of mental entities with W2, and the class of abstract entities with W3. Time is not a physical entity in the sense in which stones and rivers are: it does not belong to the physical world (W1). Time is not a specific mental state (W2), such as the pain I feel in my left knee and my love for Esmeralda, but it is a creation of the human mind and it belongs to the space of abstract entities (W3). Time is an abstract means by which we measure the flow of physical reality, or more precisely, the amount and intensity of its change. All measures are products of the human mind; therefore, they are abstract entities. The discourse about the flow of time points at the ephemerality of all physical phenomena, but ephemerality is immanent to physical phenomena: time does not cause it; time only describes it. In sum, (1) stones, (2) my feelings, and (3) time all exist and can be called "real", but they dwell in different spaces of existence and reality.

Let us see the second column of our matrix of basic concepts. In a discourse about knowledge, it is necessary to differentiate (1) *the structure of reality in itself*, (2) *a formal (mathematical) description* of a class of phenomena, and (3) *the interpretation of a formal description*, which explains what a formal description actually means.

The structure of reality in itself is unreachable, by definition, and it seems inconceivable. It is not possible to know reality in itself, because knowledge is a *relational* category: knowledge is a "joint product" of the observed object and of the capacities, means, and aims of the subject that observes the object. Our knowledge is never absolute or "pure": we always shape what we see; we do it by our capacities, means, and aims.

By the formal description of phenomena we mean the description by means of formulas. A formula expresses the relationship between certain attributes of a class of phenomena: it defines the behaviour of an attribute (of a phenomenon of that class) in terms of the behaviour of some other attributes of that phenomenon. A formula tells how a change of certain attributes of a phenomenon influences (causes) the change of some other attribute of that phenomenon.

Formulas receive and produce quantities (numbers): they function like machines which receive quantities as inputs and produce quantities as outputs. It is not possible to prove that a formula is correct; its correctness can be tested and corroborated by experiments and observations, but it is always possible that some future case will show that the formula is not correct. If that happens, we must make some change in the formula or in the scope of its application.

Formulas are mute: they do not say what they, their inputs and outputs *mean*. Hence, a formula is normally accompanied by an explanation which says what the formula actually means or tells in ordinary language. Such an explanation is the interpretation of the formula and a partial interpretation of the theory of which the formula is a part. Formulas can be empirically tested, but interpretations cannot be tested in such a direct way. A formula can be interpreted in different ways, and *a correct formula can be interpreted in a wrong way*. Physicists test formulas, and when the results of tests confirm a formula, they normally take this as the confirmation of their *interpretation* of the formula, too. But this is not a logically valid step and it often leads to a wrong conclusion. Let me mention a notorious example in this regard.

Physics knows various kinds of particles, which have small decay half-times or half-lives. A half-life is the time in which half of the particles decay; in the next half-life, half of the remaining particles decay, and so on. Some particles have a very small rest mass, so they can be accelerated to a very high speed, close to the speed of light. By means of a formula called Lorentz transformation, we can compute the increase of half-life of the particles which move with the speed *v* relative to the earth. The results computed by this formula have been empirically confirmed many times, so this formula is considered correct. However, here ends the factual discourse; the *explanation* of these facts is a matter of interpretation. My interpretation of the increase of half-life of particles runs like this. Particles are processes (they decay); with the increase of speed, *processes* evolve more slowly. On the other hand, relativists interpret the same empirical fact in the following way: with the increase of speed with which an entity moves, *its time* flows slower. I argue that the discourse about the slowing down of time is inconsistent, which means that it is wrong. In any case, relativists cannot use the correctness of *formulas* as the proof of the correctness of their *interpretation* of these formulas. But that is what they always do.

Let us see the third and last column of our matrix of basic concepts. We can speak about reality at three basic levels: (1) *ontological*, (2) *epistemological*, and (3) *logical*. Ontology aims to establish what are the basic ingredients that reality is made of, and what are the features of these ingredients. Since it is not possible to know reality in itself, ontology tries to find a set of basic concepts (categories) by means of which (and those derivable from them) we can *describe* the reality we experience, in the most suitable way. Such *descriptive ontology* is hypothetical: it analyses conceptual systems, points at their drawbacks and seeks better solutions.

Epistemology deals with the nature, origin, and limits of human knowledge. Epistemology aims to establish the principles and methods on the basis of which we can decide which claims ought to be considered *knowledge* and which ought not. The core issue of epistemology regards the problem of justification (verification) of claims. It is difficult (or impossible) to tell whether a justification is really valid, but we do not need to deal with this issue here.

Logic deals with the issues of validity of inference and of consistency of discourse. Logic seeks and finds its basic principles in the structural features of the human language, reasoning, and understanding. What makes logic particularly important here, is that it sets the threshold below which a discourse necessarily contains false claims (statements) and it loses a clear meaning: this threshold is consistency. A discourse (theory) is consistent if and only if it is not inconsistent; a discourse is inconsistent if and only if it logically implies a statement *p* and its negation *not(p)*. In other words, an inconsistent discourse claims and negates the same thing, so that it surely contains a statement that is not true, and it cannot be a correct description of anything. An inconsistent discourse does not express a clear meaning: it is not possible to tell what an inconsistent theory says, because it negates what it claims.

Discourses in physics often show a lack of understanding of logical concepts. For example, speaking about the issue of travelling through time, Paul Davies says that "the laws of the universe must by definition describe a consistent reality" (p. 249). A better way to explain the relationship between laws (discourse) and physical reality could run like this. Physical reality *is* such as it is; it makes no sense to call it either consistent or inconsistent: consistency regards conceptual systems (theories, discourse), not material systems. A theory as a set of claims (formulas, laws) must be logically consistent to be understandable (even if wrong) and to have a chance to describe a reality in a correct way. An inconsistent theory necessarily implies factually wrong claims, and it does not express a clear content (meaning).

**3. The origin of time**

For a discourse about time, we need concepts of *past* and *future*. For the past and future to exist, we need a *now* which separates them and in relation to which they *are* the past and the future. "Now" is a *state of mind* and it exists only for the conscious mind. In the inanimate physical world there is no "now", nor can physical theories express such a state. Hence, there is also "no future or past, and no arrow or flow of time in the inanimate world" (Fraser, 242). The inanimate physical world is *now-less*, and consequently *time-less*. Time is not a cosmic phenomenon, ontologically independent of the mind. Events in the inanimate physical world can be considered "ordered" or "directed" in some way, but there is no basis for considering this order *temporal* in itself. "Now" is a subjective state, which transcends the language of physics. Physics can describe only a now-less reality, which means a time-less reality.

Time is born with the appearance of consciousness. Without conscious beings, the events on the earth, life included, would evolve as a *reality in change*, as they were evolving before the conscious mind appeared on its surface. There would not be *time* in such a world, because it is not possible to speak about *time* as something independent of the conscious mind. This is a conceptual issue, not empirical: clocks do not know time; only a conscious mind can know time. A lot has been said about the mind, but we do not have a substantial theory of the mind. I use the concept "fissure" as a figurative description of the relationship between the conscious mind and the physical reality from which it emerges. Each consciousness is a *fissure* on the face of physical reality, in which the existence reflects itself. Consciousness is not a new kind of substance, but it brings a radically new *quality* into existence; let it then be called a fissure on physical reality. "Now" does not move; physical reality changes ("flows") in front of the conscious mind ("fissure") that observes it.

There are claims that the passage of time is necessary for changes to take place. The relationship between time and change seems problematic, but this is a consequence of the wrong approach to this issue. It is wrong to start from the position that events take place "in time" and that change "needs time" to take place. Change is *immanent* to physical reality; change is also the basic element of the human perception and understanding of that reality. Physical reality is a process of ceaseless becoming and vanishing. Change is ontologically and epistemologically prior to time: we perceive change, not time. If there were no change, nobody would speak about time: people would have not *invented* it. Time is an abstract means by which the mind describes its perception and understanding of reality as a process and change. Time is a means of discourse; it does not exist beyond the human mind and language, except in the realm of abstract entities created by the human mind and language.

The basic assumption of physics is that "there is a real world out there that we can make sense of. And that world includes time", says Davies (p. 43). We do assume that the world is out there, although we cannot prove that this is really so. But the world out there does *not* include time. Time is an element of the conceptual system by means of which the mind tries to make sense of what is out there; but time is not out there. *Change* is out there, but not time. Change is immanent to physical reality: it simply *is*. There is nothing "inferential" in the perception of change: "it is simply given in the experience", says Robin Le Poidevin (p. 87). There are claims that "we perceive time through motion" (Davies, 29), but such claims are wrong. We do not perceive time through anything, because there is no time in the physical world to be perceived. We perceive *change;* on the basis of this experience, we created time as one dimension of the space of discourse about reality, on which we express our experience of change.

We cannot deal with the history of ideas about time here; let us only mention that Aristotle, Augustine, Newton, and Leibniz spoke about time in the ways which have essential similarities with the view we put forward here, but the discourse of each of them differs from our view in a certain element (cf. Westphal and Levenson). Aristotle essentially reduces time to "a measure of motion", but his discourse about this issue seems circular and he admits that this issue is difficult. Augustine argued that time is essentially related to the mind, and that it is not an ingredient of the inanimate physical reality. Newton introduced "mathematical" time which "flows equably" and is independent of everything else. Newton's mathematical time is similar to our time (as an abstract entity), but his time "flows", while our time does not: our time is the bank (coordinate) in relation to which physical reality flows. Leibniz argued that Newton's mathematical time does not exist; the only thing that exists and that we can know is the "order of succession" of events: there are no "moments" apart from events, he argued. The two masters quarrelled about this issue which can be solved easily. We can take that Newton's mathematical time is the *abstract image* of Leibniz's order of succession in the case of an absolutely stable cyclic process. On the other hand, to speak about the order of events, we need an *idea* of order, such as the one given by Newton's description of mathematical time. In sum, the two discourses *need* each other and together they lead to the same abstract entity which we call time.

**4. Does physics know time?**

It is considered that physics knows best what time is and how it looks like, because time plays a prominent role in its theories. However, physics consists of many theories; each of these theories describes a class of phenomena observed at a certain level of observation and from a certain point of view. For example, the theory of relativity describes physical reality at the level of massive bodies, while the quantum theory describes the same reality at the subatomic level of description. Both these theories are successful at the operative level. But "these two theories are known to be inconsistent with each other", so that "they cannot both be correct", says Stephen Hawking (p. 12).

Different theories in physics often give different images of time, so that it is not possible to tell what "time" actually *means* in physics, beyond its technical role in a specific theory. In any case, physics is not in position to tell us *what time is* and *how it flows*, because different physical theories describe time in different ways which are not mutually compatible or even internally consistent. Describing reality at the subatomic (quantum) level, Thomas Fraser speaks about a "nonflowing, undirected, now-less time" (p. 106). Observed at the level of quantum theory, "the universe has no well-defined time at all", says Paul Davies (p. 181); Dieter Zeh speaks about the "timeless quantum world" (p. 197). Such discourse opens some essential question which we cannot deal with here. For example, when and how does time *appear* in the physical world if it does not exist at the subatomic level?

Theory of relativity was developed in two stages; the result of the first stage is called special theory of relativity (STR); the result of the second stage is called general theory of relativity (GTR). Relativists proclaimed Newton's universal space and time, which are equal everywhere and for everybody, "fundamentally flawed" (Davies, 32), but this makes their discourse about motion problematic. In Newton's physics, every motion can be registered in relation to the abstract space which is at absolute rest; speed is also determined in relation to the fixed space. In the theory of relativity, the motion of an entity is observed in relation to other entity. However, without Newton's absolute space, it is not possible to tell *who* actually moves. Because A cannot move in relation to B, without making B move with the same speed in relation to A. This makes the discourse of STR controversial, but the idea of the relativity of time has been so *charming* that many physicists refuse to see the problem we mentioned. We discussed this notorious problem elsewhere (Radovan 2011; Radovan 2013), so that we will nod deal with it here. Let us see the narrative related to GTR, which brings even more curious things than STR.

GTR adopted a non-Euclidean geometry and a four-dimensional space of discourse about physical reality, in which space and time are joined into *spacetime*. The existence of an entity is described by a curve in the four-dimensional space, called *world line*. Points of such curve can be labelled by their three spatial parameters and by the successive values of the parameter *t* associated with the clock carried by the entity. In the spacetime, events do not happen: they always *are;* the illusion of happening (change) is created by the movement of the observer in that realm of existence. "The objective world simply is, it does not happen", runs the statement attributed to Hermann Weyl. Only to the "gaze" of a consciousness, "crawling upward" along the world line of the body (from which it emerges), "does a section of the world come to life as a fleeting image in space which continuously changes in time" (in Whitrow, 348).

GTR has been accompanied (interpreted) by the *block universe* view of existence, which takes that reality is a four-dimensional space that contains all entities and events. The block universe comprises all "moments" (or versions) of every entity that *is* (had been or will be). Time is one of the four dimensions of the block universe, but it does not flow *in* this universe, and there is no "privileged now" in such a universe (Price, 13). Every entity and event is always there (in the spacetime), regardless when it enters into the experience of an observer. Happening is an illusion created by the consciousness that "crawls" along the word line of the body from which it emerges. However, it seems that the discourse about "crawling" contradicts the claim that everything "is" and nothing "happens": is not this crawling a happening? The block universe view has a mystical charm and the appeal of a fairy tale, but it lacks a clarity and precision of scientific discourse. Contrary to Weyl, we claim: *the objective world is not: it* *is constantly becoming and vanishing*. Existence is a process: nothing is; everything is becoming and vanishing, including us, the observers of this ceaseless play.

In the block universe, all entities and events which we consider past, present or future, are equally real, says Davies; all people, past, present and future, are "there" (p. 260). "The entire universe exists as a single block with no parts, so no part of it either comes into existence or goes out of existence", explains Raju (p. 257). Such claims tell nothing to me: I do not understand them. "The most natural way of thinking of time, in the context of contemporary physics and cosmology, is to regard it as one dimension of a multidimensional space-time manifold", says Lockwood. And "our best theories" support the block universe interpretation of GTR (p. 249). However, Lockwood admits that the block universe view "remains a controversial view, which may conceivably turn out to be mistaken" (p. 249). Let me remind you that this view is an interpretation, which cannot be empirically tested in a direct way. This interpretation is vague, so that it is not clear *in what way* could it be tested and shown either correct or mistaken at the level of empirical facts. The block universe view is poetry rather than physics.

Newton introduced absolute space and time as the conceptual means that facilitate a precise discourse about physical reality. His space and time serve for "keeping track of motion mathematically", but they do not "*do* anything", says Davies. On the other hand, Einstein "restored time ... at the heart of nature, as an integral part of the physical world" (p. 17). In the theory of relativity, "space and time ... play a full and active role in the great drama of nature" (p. 16). This sounds exciting, but we consider wrong to assume that time is part of the physical world and that it *does* anything. Relativists argue that "space and time ... are *physical* things, mutable and malleable, and, no less than matter, subject to physical law" (Davies, 16). These quotations put forward the essence of what we consider wrong in the discourse about GTR and its block universe interpretation. Space and time do not play any role in the drama of nature; they are not physical things, mutable and malleable, and they are not subject to physical laws. Space and time are *abstract* entities (categories), created by the human mind, and they exist only in the realm of abstract entities (W3). Space and time facilitate precise *discourse* about physical reality; they are *not part* of physical reality.

Lockwood stresses that the theory of relativity made "space and time, as aspects of space-time, active participants in physical interactions" (p. 79). The old conception of space as "a merely passive arena" in which forces play their games, and of time as "merely the road along which the march of history proceeds", has given way to "an actively involved space-time", which "feels" the matter and energy, and "kicks back" accordingly (p. 79). This sounds exciting, but it tells nothing, because this is a play with metaphors, not a concrete discourse.

Relativists routinely say that in Newtonian discourse space and time form the "arena" or the "stage" in/on which physical events unfold. But this is not correct. In Newton's physics, space and time are the abstract means by which we describe physical reality as a process that unfolds in nothingness (because there is nothing beyond it). Time and space are not the stage: physical reality does not need any stage; *people need concepts* to be able to describe what they perceive and experience. By mixing basic categories - the physical and the abstract (language) - physics can produce nice metaphors which do not explain anything.

## 5. Concluding remarks

Time is a conceptual rather than an empirical issue: it is a matter of metaphysics and logic rather than a matter of physics and testing. We must *define* time and its features rather than discover them in the physical world, and we must do this in a consistent way. Physical reality is a process: it *is* as a ceaseless becoming and vanishing. Change is immanent to physical reality; the experience of change is the source of the idea and of the concept of time. Time is not a part of physical reality and it does not flow, either absolutely or relatively, either forward or backward. Time is the abstract bank in relation to which physical reality flows. Time belongs to language: it is one of the basic concepts in terms of which we describe physical reality and the way it changes.

The fact that speed and gravity slow down processes ("clocks") does not mean that they slow down time. The former is a verifiable fact; the latter is an inconsistent interpretation. Davies considers Einstein's discourse about space and time "a monumental first step" in the right direction, but he admits that this revolution remained "frustratingly unfinished" (p. 17). We argue that this revolution is incompletable because its discourse about the relativity of time is inconsistent. The assumption that time and space are relative and that they are "physical things" and "malleable" leads to contradictions. It is necessary to differentiate physical reality and abstract entities (language) by means of which we describe the reality.

A lot of things have been said about the beginning of time. We cannot elaborate this issue here; let us just mention that time should be considered infinite for semantic reasons. The concept of beginning can have a meaning only *in* time, so that we cannot speak about the beginning of time itself in a coherent way. Time did not begin with the Big Bang, because the Big Bang was not the absolute beginning. Every "bang" is a process, and every process is caused by another process. It is not possible to speak about the absolute beginning because *something* (existence) cannot emerge out of *nothing*. No cognitively relevant explanation about the origin of existence can exist.

Physics will abandon time one day and replace it with the *intensity of change*, but people will keep the plain old time for their reflections on existence and ephemerality. Because although time does not pass, everything else does. People will always speak about the flow of time which caries everything away, but this is only a figurative way of saying that existence is a process in which things are ceaselessly becoming and vanishing.

+